\documentclass[preprint, 3p, sort&compress]{elsarticle}
\usepackage{dcolumn}
\usepackage{booktabs,tabularx} 
\usepackage{booktabs}
\usepackage{textcomp}
\usepackage{amssymb}
\usepackage{amsthm}
\usepackage[mathlines]{lineno}
\usepackage{graphicx}
\usepackage{units}
\usepackage{url}
\usepackage{amsmath}
\usepackage{amsfonts}
\usepackage{bm}
\usepackage{textcomp}
\usepackage{subfigure}
\usepackage{multicol}
\usepackage{multirow}
\usepackage{verbatim}
\usepackage{rotating}
\usepackage[colorlinks,linkcolor=black,citecolor=red]{hyperref}
\usepackage{float}
\usepackage{epsfig}
\usepackage{dcolumn}
\usepackage{bm}
\usepackage{color}
\usepackage{pstricks}
\usepackage{pst-node}
\usepackage{times}
\usepackage{indentfirst}
\usepackage[english]{babel}
\usepackage{epstopdf}
\addto{\captionsenglish}{%

}
\makeatother
\usepackage[font=footnotesize]{caption}
\usepackage{overpic}

\biboptions{numbers,sort&compress}
\begin{document}
\markboth{Zhi Gao, Tianjin (CN)}{ }
\begin{frontmatter}
\title{BSM searches at BESIII
\tnotetext[invit]{Talk given at QCD25 - 40th anniversary of the QCD-Montpellier Conference. }
\author{Zhi Gao}
\address{NanKai University, Tianjin 300071, China}
\ead{zhi-gao@mail.nankai.edu.cn}
}
\date{\today}
\begin{abstract}
The BESIII experiment is a symmetric $e^+e^-$ collider experiment operating at center-of-mass energy from 2.0 to 4.95 GeV. With the world’s largest threshold production data set of $J/\psi$ (10 billion), $\psi(3686)$ (2.7 billion), $7.9~ \rm fb^{-1}$ $D$ meson pairs from $\psi(3770)$ decay, and $7.33~ \rm fb^{-1}$ $D_s D_s^*$ events from 4.128 to 4.226 GeV, we are able to probe for new physics through precision tests of the SM, searches for exotic low-mass particles, and investigations of forbidden or rare decay processes. In this talk, we report recent studies on Beyond the Standard Model physics conducted by the BESIII collaboration, including several specific searches for axion-like particles, dark photons, and QCD axions, search for invisible decay of $K_S^0$. Meanwhile, a series of rare cham decay processes, including searches for lepton/baryon number violation, FCNC processes and charmonium weak decay processes, can also be probed to search for new physics at BESIII.

\end{abstract}
\end{frontmatter}


\section{Introduction}
The standard model (SM) can explain nearly all experimental results in particle physics, yet it fails to account for the dark matter in the universe, nor the mechanism to produce the dominance of matter over antimatter in the universe. The neutrino mass, the number of particle generations, and the gravity also lie outside its scope.

\begin{itemize}
  \item The universe is dominated by dark energy and dark matter (DM), where the amount of non-baryonic dark matter is about 1/4 of the critical density of the universe. DM problem requires new particles. Direct detection experiments like XENONnT~\cite{XENONnT}, and DAMA~\cite{DAMA} experiments search for DM particles, though conclusive evidence remains difficult to replicate. 
  \item To explain the matter-antimatter asymmetry in universe, many baryogenesis theoretical models are proposed. The theoretical models of particle physics can be constrained using astrophysical observations and can also be  constrained by collider experiments.
  \item The neutral particles with tiny mass, neutrinos, play an important role in generating the baryon asymmetry  and contributing to the dark matter. Some solar neutrino experiments~\cite{solarneutrino1,solarneutrino2,solarneutrino3} as well as laboratory experiments including neutrinoless double beta decay experiments~\cite{0nu2beta1,0nu2beta2,0nu2beta3} have given some constrains on the mixing angles and Majorana effective mass.
\end{itemize}

Collider experiments like LHC also contribute to the investigations of BSM physics \cite{LHCI,LHCII,LHCIII,LHCIV,LHCV}. By studying physics in the tau-charm energy region with massive charmonium data samples, BESIII~\cite{BESIII} conducts precision SM tests and sensitive searches for low-mass exotic states and rare processes that could be portals to new physics. In this talk, we review some of recent highlights of BSM searches at BESIII experiment.

\section{BSM particle searches}
Although the heavy exotic resonances  are beyond the energy region of BESIII, we have carried out some searches for low mass exotic particles.
 For instance, the CP-odd light Higgs in Higgs sector extended by next-to-minimal supersymmetric standard model \cite{NMSSM}, as discussed in~\cite{CPoddHiggs}. Additionally, searches for dark sector particles, which represent a collection of particles that are not charged directly under the SM strong, weak, or electromagnetic forces, have been carried out at BESIII. theses searches of dark sector particles can be  categorized  into three types: searches for visible processes~\cite{visibleI,visibleIII}, searches for fully invisible processes~\cite{invisibleI,invisibleII,invisibleIII}, and searches for invisible signature~\cite{invisiblesignI,invisiblesignII,invisiblesignIII,invisiblesignIV}. In this talk, we focus on several searches for dark sector particles including axion-like particles (ALP), dark photons, and QCD axions.

\begin{figure}[H]
\begin{center}
\includegraphics[width=7cm]{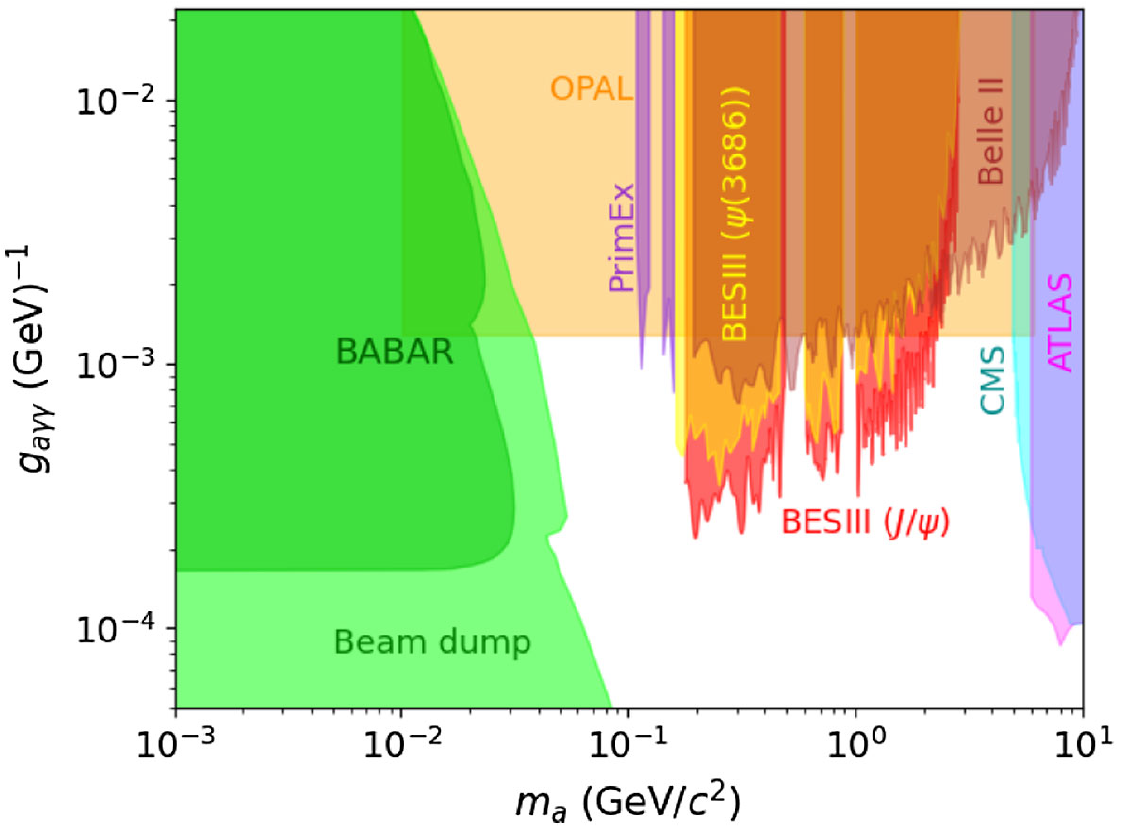}
\vspace{-0.2cm}
\caption{The 95\% confidence level (C.L.) upper limits on the coupling of an ALP to a photon pair~\cite{invisiblesignIII}.}\label{fig::cpagamma}
\end{center}
\vspace{-0.2cm}
\end{figure}

ALPs are generalization of QCD axions, appearing in theories like string theory and extended Higgs models. For very light ALPs, the di-photon channel is the only SM decay channel.
In 2023, 2024, BESIII have searched for ALPs in radiative $J/\psi$ decays via $J/\psi\to a\gamma, a\to \gamma\gamma$~\cite{invisiblesignII,invisiblesignIII}. In the analyses based on 2.7 billion $\psi(2S)$ events and 10 billion $J/\psi$ events, the upper limits on the coupling constant of an ALP to a photon are set as a function of the mass of ALP. And the limits obtained at BESIII are the most stringent in the mass region from 0.18 GeV to 2.85 GeV, as shown in Fig. \ref{fig::cpagamma}~\cite{invisiblesignIII}.

The dark photon, served as a portal between SM particles and dark sector, can be described by a minimum extension of the SM requiring an extra Abelian gauge group, which will cause the associated spin-one boson.  The massless dark photon, with the symmetry remaining unbroken has no interaction with the SM particle within the dimension-four operator. A dimension-six effective field theory (EFT) operator defined as Eq. \ref{eq::D6op}~\cite{D6Operator} can couple the dark photon with two quarks or two charged leptons.
\begin{equation}\label{eq::D6op}
\begin{aligned}
    \mathcal{L}_{\rm NP}=&\frac{1}{\Lambda_{\rm NP}^2}(C_{jk}^{U}\bar{q}_{j}\sigma^{\mu\nu}u_{k}\tilde{H}+C_{jk}^{D}\bar{q}_{j}\sigma^{\mu\nu}d_{k}H  \\
     +& C_{jk}^{L}\bar{l}_{j}\sigma^{\mu\nu}e_{k}H+h.c.)F_{\mu\nu}',
\end{aligned}
\end{equation}
Here $\mathcal{L}_{\rm NP}$ means an effective mass, indicating the new physics energy scale. The first item of the dimension-six operator will cause the $cu \gamma'$ coupling, which is the flavor-changing neutral current (FCNC) of charm quark. More details about the dimension-six EFT operator can be found in Ref.~\cite{FCNCI}.

In SM, FCNC processes are strongly suppressed by the Glashow-Iliopoulos-Maiani (GIM) mechanism and the branching fraction (BF) of the FCNC processes of the charm quark predicted by standard model would not exceed the level of $10^{-9}$. But for $cu \gamma'$ coupling, its FCNC process is originated from the new physics energy scale, and the BF can reach to the order of $10^{-7}\sim 10^{-5}$~\cite{FCNCI}. BESIII collaboration has searched for the massless dark photon with $D^{0}\to\omega\gamma'$ and $D^{0}\to\gamma\gamma'$~\cite{FCNCI} in 2024. The constraint on the parameter related to the new physics energy scale is set to be $|\mathbb{C}|^2 + |\mathbb{C}_5|^2<8.2\times 10^{-17}~\rm GeV^{-2}$~\cite{FCNCI}, which is the first exploring of the DM and vacuum stability allowed space.

\begin{figure}[H]
\begin{center}
\includegraphics[width=7cm]{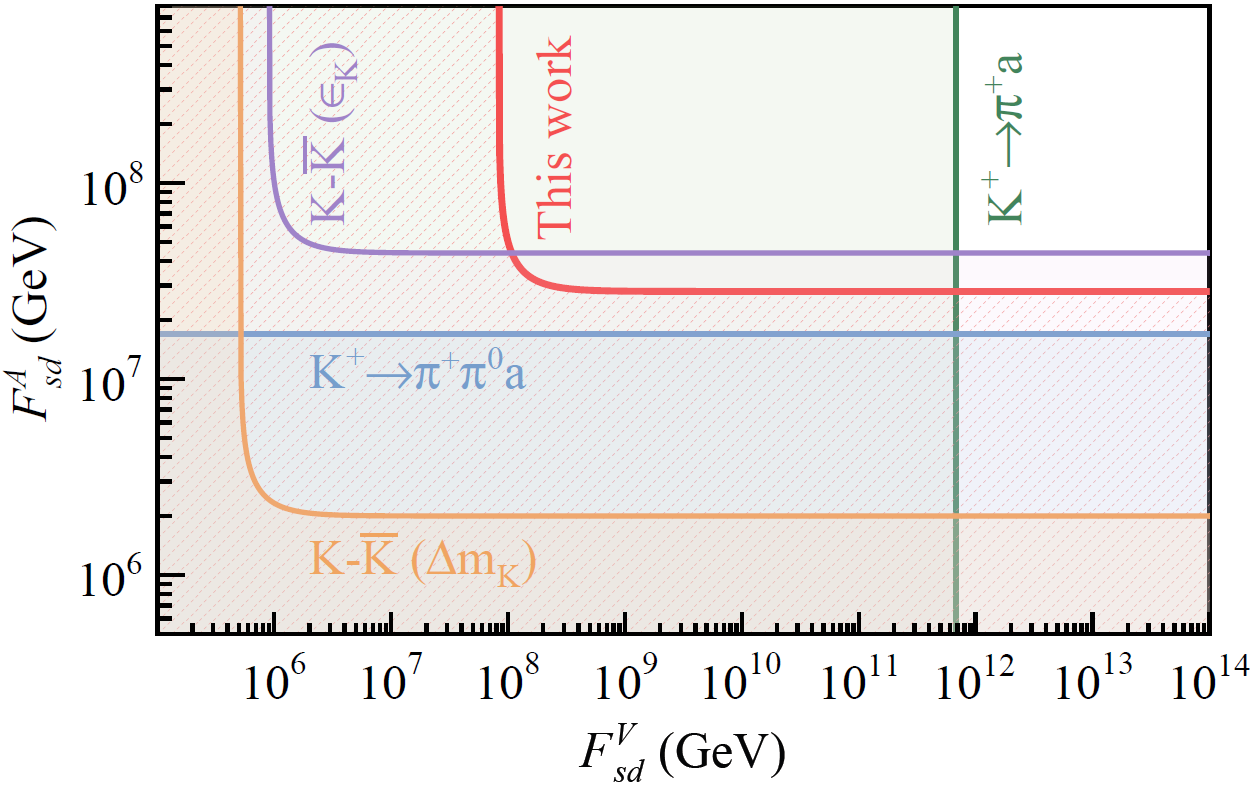}
\vspace{-0.2cm}
\caption{The 90\% CL exclusion limits of $s\to d$ axion-fermion effective decay constant~\cite{invdecayIV}.}\label{fig::FAsd}
\end{center}
\vspace{-0.2cm}
\end{figure}

FCNC process can also be introduced by a QCD axion which is an excellent cold dark matter candidate. The mass of QCD axion is inversely proportional to its symmetry breaking scale and is lower than one eV. The low mass and long lifetime of the QCD axion make it possible to search in the invisible process $\Sigma^{+}\to p+invisible$~\cite{invdecayIV}.  The lower bound of the effective decay constants for the axial coupling term $F_{sd}^{A}$ is set to be $2.8\times 10^{7}~\rm GeV$ as illustrate on Fig.~\ref{fig::FAsd}~\cite{invdecayIV}.

\section{Invisible decays}

Studies of invisible decays are important for SM extensions. Due to the weak coupling between light DM and SM particles, it is challenging to detect them using traditional observational methods. At collider experiments like BESIII, invisible signatures potentially associated with light DM can be searched for by imposing constraints from four-momentum conservation.
BESIII have carried out some invisible searches in sub GeV regions~\cite{invdecayI,invdecayII,invdecayIII,invdecayIV}.

The search for invisible decay of $K_{S}^{0}$~\cite{invdecayI} has been conducted in 2025. Investigations of $J/\psi\to\phi K_{S}^{0} K_{S}^{0}$ offers a sensitive test of the SM and the mirror matter model. In addition, under Wigner-Weisskopf approximation and using Bell-Steinberger relation, the CPT-violating parameters can be connected with the physical kaon decay amplitudes. This allows the $K_{S}^{0}$ invisible decay search to offer a direct experimental basis to perform CPT tests in the $K^0-\bar{K}^0$ system, without assuming that there is no contribution from invisible decay modes.

The discrepancy between experimental measurement of the muon anomalous magnetic moment and the SM prediction may also indicate BSM physics. A muonphilic vector boson $X_1$ introduced within an anomaly-free gauged model~\cite{X1} can contribute to the muon anomalous magnetic moment. For $X_1$ masses below $2m_{\mu}$, BESIII experiment can search for its invisible decays via $J/\psi$ decays~\cite{invdecayIV}.

\section{Baryon/Lepton number violation processes}

Lepton and baryon number violation (LNV/BNV) are connected with the Baryon asymmetry through some baryogenesis models like leptogenesis models~\cite{lepto}. Many experiments are searching for neutron-antineutron oscillations and proton decay. BESIII experiments enrich these studies by investigating LNV/BNV processes, for instance the search for $\Lambda-\bar{\Lambda}$ oscillation in the decay $J/\psi\to pK^-\Lambda+c.c.$~\cite{BNV} conducted in 2023. In this analysis, we calculate the time-integrated $\Lambda-\bar{\Lambda}$ oscillation rate $P_{\Lambda}$ by the ratio of BFs of $J/\psi \to p K^-\Lambda$ and $J/\psi \to p K^- \bar{\Lambda}$ with the assumption of no CP violation, and then the upper limit on the oscillation parameter is set to be $3.8 \times 10^{-18}$ at 90\% C.L.

\section{Other rare charm decays}

More FCNC processes  in charmed meson and charmonium decays have been carried out at BESIII recently. No significant signals for BSM physics are found and the upper limits on the BFs of various FCNC processes at 90\% C.L. are summarized in Table~\ref{tab:FCNC}.

\begin{table}[H]
\begin{center}
\setlength{\tabcolsep}{0.1pc}
  \caption{\footnotesize
Upper limits on the BFs ($\mathcal{B}^{\rm ul}$) of various FCNC processes at 90\% C.L. searched at BESIII.}
    {\footnotesize
    \begin{tabular}{cccc}
\hline
Decay & Data & $\mathcal{B}^{\rm ul}$ at 90\% C.L. &  Ref. \\
\hline
 $D_s^+\to \pi^+\pi^0 e^+e^-$&$7.3 {\rm fb^{-1}}@4.128\sim 4.226~\rm GeV$&$7.0\times 10^{-5}$& \cite{FCNCII}\\
 $D_s^+\to K^+\pi^0 e^+e^-$&$7.3 {\rm fb^{-1}}@4.128\sim 4.226~\rm GeV$&$7.1\times 10^{-5}$& \cite{FCNCII}\\
 $D_s^+\to K_S\pi^+ e^+e^-$&$7.3 {\rm fb^{-1}}@4.128\sim 4.226~\rm GeV$&$8.1\times 10^{-5}$& \cite{FCNCII}\\
  $J/\psi\to D^{0}\mu^+\mu^-$&$1.0087\times 10^{10}@3.097~\rm GeV$&$1.1\times 10^{-7}$&\cite{FCNCIII}\\
\hline\hline
\end{tabular}
}
\label{tab:FCNC}
\end{center}
\end{table}

Rare weak decays of $J/\psi$ to a $D$ or $D_s$ meson accompanied with a light meson are highly suppressed in SM, with BFs predicted to be the order of $10^{-8}$~\cite{SMweakdecay}. While in some BSM models~\cite{weakdecaymodelsi,weakdecaymodelsii,weakdecaymodelsiii}, BFs are predicted up to the order of $10^{-5}$. Table~\ref{tab:charmweak} summarized the results of upper limits on the BFs of various charmonium weak decays searched at BESIII.
\begin{table}[H]
\begin{center}
\setlength{\tabcolsep}{0.1pc}
  \caption{\footnotesize
Upper limits on the BFs of various charmonium weak decays at 90\% C.L. searched at BESIII.}
    {\footnotesize
    \begin{tabular}{cccc}
\hline
Decay & Data & $\mathcal{B}^{\rm ul}$ at 90\% C.L. &  Ref. \\
\hline
$J/\psi\to D^-\rho^+$ &$1.0087\times 10^{10}@3.097~\rm GeV$&$6.0\times 10^{-7}$&\cite{charmweakdecayi}\\
$J/\psi\to D^-\pi^+$&$1.0087\times 10^{10}@3.097~\rm GeV$&$7.0\times 10^{-8}$&\cite{charmweakdecayi}\\
$J/\psi\to \bar{D}^0\rho^0$&$1.0087\times 10^{10}@3.097~\rm GeV$&$5.2\times 10^{-7}$&\cite{charmweakdecayi}\\
$J/\psi\to \bar{D}^0\eta$&$1.0087\times 10^{10}@3.097~\rm GeV$&$6.8\times 10^{-7}$&\cite{charmweakdecayi}\\
$J/\psi\to \bar{D}^0\pi^0$&$1.0087\times 10^{10}@3.097~\rm GeV$&$4.7\times 10^{-7}$&\cite{charmweakdecayi}\\
$J/\psi\to D_s^-\rho^+$&$1.0087\times 10^{10}@3.097~\rm GeV$&$8.0\times 10^{-7}$&\cite{charmweakdecayii}\\
$J/\psi\to D_s^-\pi^+$&$1.0087\times 10^{10}@3.097~\rm GeV$&$4.1\times 10^{-7}$&\cite{charmweakdecayii}\\
\hline\hline
\end{tabular}
}
\label{tab:charmweak}
\end{center}
\end{table}

\section{Summary}
BESIII experiment contributes significantly to BSM physics study by accumulating largest datasets of $J/\psi$, $\psi(3770)$ decays, and $D_s D_s^*$ events, enabling high-sensitivity searches for low-mass BSM particles and rare or forbidden decays. BESIII recently collected $12.3~\rm fb^{-1}$ events at $3.773~\rm GeV$, and a data-taking is ongoing for scan beam energy regions and $\psi(2S)$ production. BESIII offers strong potential for future BSM physics searches as the accumulated larger datasets undergo analysis.

\section*{Acknowledgement}

The author would like to thank all the colleagues in BESIII Collaboration for their great effort to make the mentioned results available, and of course thank the organiser of QCD2025 conference to make the extraordinary event happen. The talk is supported by National Natural Science Foundation of China (Grants No. 12035009).



\begin{thebibliography}{00}

\bibitem{XENONnT}\href{https://doi.org/10.1103/PhysRevLett.131.041003}{ E. Aprile {\it et al}. (XENON Collaboration), Phys. Rev. Lett. {\bf 131} (2023) 041003}.
\bibitem{DAMA}
\href{https://doi.org/10.1140/epjc/s10052-008-0662-y}{R. Bernabei, P. Belli, F. Cappella, R. Cerulli, C. J. Dai, A. d’Angelo, H. L. He, A. Incicchitti, H. H. Kuang, J. M. Ma, F. Montecchia, F. Nozzoli, D. Prosperi, X. D. Sheng, and Z. P. Ye, Eur. Phys. J. C {\bf 56} (2008) 333}.
\bibitem{solarneutrino1}\href{https://doi.org/10.1103/PhysRevLett.128.091803}{M. Agostini {\it et al}. (Borexino Collaboration), Phys. Rev. Lett. {\bf 128} (2022) 091803}.
\bibitem{solarneutrino2}\href{https://doi.org/10.1103/PhysRevC.88.025501}{B. Aharmim {\it et al}. (SNO Collaboration), Phys. Rev. C {\bf 88} (2013) 025501}.
\bibitem{solarneutrino3}\href{https://doi.org/10.1103/PhysRevD.99.012012}{M. Anderson {\it et al}. (SNO+ Collaboration), Phys. Rev. D {\bf 99} (2019) 012012}.
\bibitem{0nu2beta1}\href{https://doi.org/10.1103/RevModPhys.80.481}{F. T. Avignone, S. R. Elliott, and J. Engel, Rev. Mod. Phys. {\bf 80} (2008) 481}.
\bibitem{0nu2beta2}\href{https://doi.org/10.1007/JHEP09(2023)190}{P. Novella et al. (NEXT Collaboration), JHEP {\bf 09} (2023) 190}.
\bibitem{0nu2beta3}\href{https://doi.org/10.1103/PhysRevLett.131.152501}{I. J. Arnquist et al. (MAJORNANA Collaboration), Phys. Rev. Lett. {\bf 131} (2023) 152501}.
\bibitem{LHCI}\href{https://doi.org/10.1007/JHEP11(2023)168}{G. Aad {\it et al}. (ATLAS collaboration), JHEP {\bf 11} (2023) 168}.
\bibitem{LHCII}\href{https://doi.org/10.1007/JHEP07(2023)125}{G. Aad {\it et al}. (ATLAS collaboration), JHEP {\bf 07} (2023) 125}.
\bibitem{LHCIII}\href{https://doi.org/10.1016/j.physletb.2023.138394}{G. Aad {\it et al}. (ATLAS collaboration), Phys. Lett. B {\bf 848} (2024) 138394}.
\bibitem{LHCIV}\href{https://doi.org/10.1007/JHEP07(2021)208}{G. Aad {\it et al}. (ATLAS collaboration), JHEP {\bf 07} (2021) 208}.
\bibitem{LHCV}\href{https://doi.org/10.1140/epjc/s10052-024-12975-4}{G. Aad {\it et al}. (ATLAS collaboration), Eur. Phys. J. C {\bf 84} (2024) 818}.
\bibitem{BESIII}\href{https://doi.org/10.1016/j.nima.2009.12.050}{M. Ablikim et al. (BESIII Collaboration), Nucl. Instrum. Meth. A {\bf 614} (2010) 345}.
\bibitem{NMSSM}\href{https://doi.org/10.1016/j.physrep.2010.07.001}{U. Ellwanger, C. Hugonie, A. M. Teixeira, Phys. Rept. {\bf 496} (2010) 001}.
\bibitem{CPoddHiggs}\href{https://doi.org/10.1103/PhysRevD.105.012008}{M. Ablikim et al. (BESIII Collaboration), Phys. Rev. D {\bf 105} (2022) 012008}.
\bibitem{visibleI}\href{https://doi.org/10.1016/j.physletb.2017.09.067}{M. Ablikim et al. (BESIII Collaboration), Phys. Lett. B {\bf 774} (2017) 252}.
\bibitem{visibleIII}\href{https://doi.org/10.1103/PhysRevD.99.012013}{M. Ablikim et al. (BESIII Collaboration), Phys. Rev. D {\bf 99} (2019) 012013}.
\bibitem{invisibleI}\href{https://doi.org/10.1016/j.physletb.2023.137785}{M. Ablikim et al. (BESIII Collaboration), Phys. Lett. B {\bf 839} (2023) 137785}.
\bibitem{invisibleII}\href{https://doi.org/10.1007/JHEP10(2020)207}{M. Ablikim et al. (BESIII Collaboration), JHEP {\bf 10} (2020) 207}.
\bibitem{invisibleIII}\href{https://doi.org/10.1103/PhysRevD.105.L071101}{M. Ablikim et al. (BESIII Collaboration), Phys. Rev. D {\bf 105} (2022) L071101}.
\bibitem{invisibleIV}\href{https://doi.org/10.1103/PhysRevD.109.L031102}{M. Ablikim et al. (BESIII Collaboration), Phys. Rev. D {\bf 09} (2024) L031102}.
\bibitem{invisiblesignI}\href{https://doi.org/10.1103/PhysRevD.105.L071102}{M. Ablikim et al. (BESIII Collaboration), Phys. Rev. D {\bf 105} (2022) L071102}.
\bibitem{invisiblesignII}\href{https://doi.org/10.1016/j.physletb.2023.137698}{M. Ablikim et al. (BESIII Collaboration), Phys. Lett. B {\bf 838} (2023)137698}.
\bibitem{invisiblesignIII}\href{https://doi.org/10.1103/PhysRevD.110.L031101}{M. Ablikim et al. (BESIII Collaboration), Phys. Rev. D {\bf 110} (2024) L031101}.
\bibitem{D6Operator}\href{https://doi.org/10.1016/j.physletb.2017.02.022}{C. Chiang and P. Tseng, Phys. Lett. B {\bf 767} (2017) 289}.
\bibitem{FCNCI}\href{https://doi.org/10.1103/PhysRevD.111.L011103}{M. Ablikim et al. (BESIII Collaboration), Phys. Rev. D {\bf 111} (2025) L011103}.
\bibitem{invdecayIV}\href{https://doi.org/10.1016/j.physletb.2024.138614}{M. Ablikim et al. (BESIII Collaboration), Phys. Lett. B {\bf 852} (2024) 138614}.
\bibitem{invdecayI}\href{https://doi.org/10.1007/JHEP05(2025)092}{M. Ablikim et al. (BESIII Collaboration), JHEP {\bf 05} (2025) 092}.
\bibitem{invdecayII}\href{https://doi.org/10.48550/arXiv.2506.10316}{M. Ablikim et al. (BESIII Collaboration), arXiv:2506.10316 [hep-ex]}.
\bibitem{invdecayIII}\href{https://doi.org/10.1103/PhysRevD.105.L071101}{M. Ablikim et al. (BESIII Collaboration), Phys. Rev. D {\bf 105} (2022) L071101}.
\bibitem{X1}\href{https://doi.org/10.1103/PhysRevD.50.4571}{R. Foot, X. G. He, H. Lew, and R. R. Volkas, Phys. Rev. D {\bf 50} (1994) 4571}.
\bibitem{lepto}\href{https://doi.org/10.1016/j.physrep.2008.06.002}{S. Davidson, E. Nardi, and Y. Nir, Phys. Rep. {\bf 466} (2008) 105}.
\bibitem{BNV}\href{https://doi.org/10.1103/PhysRevLett.131.121801}{M. Ablikim et al. (BESIII Collaboration), Phys. Rev. Lett. {\bf 131} (2023) 121801}.
\bibitem{FCNCII}\href{https://doi.org/10.1103/PhysRevLett.133.121801}{M. Ablikim et al. (BESIII Collaboration), Phys. Rev. Lett. {\bf 133} (2024) 121801}.
\bibitem{FCNCIII}\href{https://doi.org/10.1007/JHEP04(2025)061}{M. Ablikim et al. (BESIII Collaboration), JHEP {\bf 04} (2025) 061}.
\bibitem{SMweakdecay}\href{https://doi.org/10.1103/PhysRevD.110.074510}{Y. Meng, J. L. Dang, C. Liu, X. Y. Tuo, H. Yan, Y. B. Yang and K. L. Zhang, Phys. Rev. D {\bf 110} (2024) 074510}.
\bibitem{weakdecaymodelsi}\href{https://doi.org/10.1016/0370-2693(94)01660-5}{C. Hill, Phys. Lett. B {\bf 345} (1995) 483}.
\bibitem{weakdecaymodelsii}\href{https://doi.org/10.1016/0370-2693(82)90262-3}{C. S. Aulakh and R. N. Mohapatra, Phys. Lett. B {\bf 119} (1982) 136}.
\bibitem{weakdecaymodelsiii}\href{https://doi.org/10.1103/PhysRevD.15.1958}{S. L. Glashow and S. Weinberg, Phys. Rev. D {\bf 15} (1977) 1958}.
\bibitem{charmweakdecayi}\href{https://doi.org/10.1103/PhysRevD.110.032020}{M. Ablikim et al. (BESIII Collaboration), Phys. Rev. D {\bf 110} (2024) 032020}.
\bibitem{charmweakdecayii}\href{https://doi.org/10.48550/arXiv.2506.09386}{M. Ablikim et al. (BESIII Collaboration), arXiv:2506.09386v1 [hep-ex]}.

\end{thebibliography}
\end{document}